\documentclass[preprint,preprintnumbers, prd, floatfix,  superscriptaddress,nofootinbib] {revtex4-1}
\usepackage{graphicx}
\usepackage{epsfig}
\usepackage{bm}
\usepackage{amssymb}
\usepackage{float}
\usepackage{amsmath}
\usepackage{dcolumn}
\usepackage{cancel}
\usepackage[colorlinks]{hyperref}
\usepackage[usenames,dvipsnames]{color}
\usepackage{color}
\usepackage{epstopdf}
\usepackage{nccbbb}
\hypersetup{
     breaklinks=true,
    pdfstartview={FitH},    colorlinks=true,       
    linkcolor=blue,          
    citecolor=red,        
    filecolor=magenta,      
    urlcolor=blue,           
    anchorcolor=green,      
    linktocpage=true
}

\begin{document}

\title{A power-law coupled three-form dark energy model}

\author{Yan-Hong Yao}
\email{Yohann@mail.nankai.edu.cn}
\author{Yang-Jie Yan}
\email{yan\_yj@mail.nankai.edu.cn}
\author{Xin-He Meng}
\email{xhm@nankai.edu.cn}

\affiliation{Department of Physics, Nankai University, Tianjin 300071, China}

\begin{abstract}
We consider a field theory model of coupled dark energy which treats dark energy as a three-form field and dark matter as a spinor field. By assuming the effective mass of dark matter as a power-law function of the three-form field and neglecting the potential term of dark energy, we obtain three solutions of the autonomous system of evolution equations, including a de Sitter attractor, a tracking solution and an approximate solution. To understand the strength of the coupling, we confront the model with the latest Type Ia Supernova (SN \uppercase\expandafter{\romannumeral1}a), Baryon Acoustic Oscillations (BAO) and Cosmic Microwave Backround (CMB) radiation observations, with the conclusion that the combination of these three databases marginalized over the present dark matter density parameter $\Omega_{m0}$ and the present three-form field $\kappa X_{0}$ gives stringent constraints on the coupling constant, $-0.017< \lambda <0.047$ ($2\sigma$ confidence level), by which we give out the model applicable parameter range.
\end{abstract}

\maketitle

\section{Introduction}
\label{intro}
According to cosmological observations, the universe has entered a stage of an accelerated expansion with a redshift smaller than 1\cite{riess1998,perlmutter1999measurements}. Since all usual types of matter with positive pressure decelerate the expansion of the universe, a sector with negative pressure named as dark energy was suggested to account for the invisible fuel that accelerates the expansion rate of the current universe\cite{Sahni2004Dark,Carroll2000The}.

The simplest cosmological model of dark energy is the so called Lambda cold dark matter ($\Lambda$CDM) model, in which vacuum energy plays the role of dark energy. Although $\Lambda$CDM model provides an excellent fit to a wide range of astronomical data so far, such model in fact is theoretical  problematic because of two cosmological constant problems, the fine tuning problem that why is the observational vacuum density  so small compared to the theoretical one, and the coincidence problem that why is the observational vacuum density coincidentally comparable with the critical density at the present epoch in the long history of the universe. In order to alleviate the latter, various of evolving and spatially homogeneous scalar field, including quintessence\cite{Caldwell1998Cosmological}, phantom\cite{Caldwell1999A}, dilatonic\cite{Piazza2004Dilatonic}, tachyon\cite{Padmanabhan2002Accelerated}and quintom\cite{Bo2006Oscillating} etc.
were suggested to take vacuum energy's place. In these models, the resolution of the coincidence problem typically leads to a fine tuning of model parameters.

Since the experimental evidences of cosmology-specific scalars particles have not been discovered yet, there is no reason to exclude the possibility of some other high form field to be dark energy. Indeed, the three-form cosmology proposed in\cite{Koivisto2009Inflation1,Koivisto2009Three} could be a good alternative to scalar cosmology, because such high form field not only respects the Friedmann-Robertson-Walker(FRW) symmetry naturally but also can accelerates the expansion rate of the current universe without a slow-roll condition. Moreover, some interesting results, e.g.three-form with simple potentials lead to models of inflation with potentially large non-Gaussian signatures \cite{mulryne2012three}, etc., about three-form cosmology are obtained.

The coincidence problem mentioned above is just that the amount of dark matter is comparable to that of dark energy in the present universe, so it is natural to consider an interaction between these two components. As was pointed out in the paper\cite{Koivisto2012Coupled}, in comparison to coupled scalar dark energy model\cite{Amendola1999Coupled}, some new features appear in coupled three-form  dark energy model, including one that the stress tensor is modified by the interaction between two dark sectors, hence it is problematic to consider a coupled three-form dark energy model in a phenomenological way\cite{Ngampitipan2011Dynamics}, and one needs to construct it in a Lagrangian formalism. Different from modeling dark matter as point particles\cite{Koivisto2012Coupled}, we follow the thread that describing the interaction between dark energy and dark matter from a fundamental field theory point of view\cite{Micheletti2009Field} and consider dark matter as a Dirac spinor field.

The contents of this paper are as follows. In section \ref{sec:1}, we present a type of Lagrangians describing the interaction between a three-form field and a Dirac spinor field in curve space-time and then derive the field equations from such Lagrangians. In section \ref{sec:2}, we consider these field equations in a FRW
space-time by assuming the effective mass of dark matter as a power-law function of the three-form field and setting the potential of dark energy to be zero. In section \ref{sec:3}, we carry out a simple likelihood analysis of the model with the use of 580 SN \uppercase\expandafter{\romannumeral1}a data points from recently released Union2.1 compilation\cite{Suzuki2012The} and BAO data from the WiggleZ Survey\cite{Chris2011The}, SDSS DR7 Galaxy sample \cite{Percival2010Baryon}and 6dF Galaxy Survey datasets\cite{Beutler2011The}, together with CMB data from WMAP7 observations\cite{jarosik2011seven}. In the last section, we make a brief conclusion with this paper.

\section{A type of field theories of three-form and Dirac spinor in curve space-time}
\label{sec:1}
This section involves some concepts that are used to include fermionic sources in the Einstein theory of gravitation and for a more detailed analysis the reader is referred to\cite{Weinberg1973Gravitation,birrell1984quantum,wald1984general,ryder1996quantum}.

A type of Lagrangians which describe the interaction between a canonical three-form field $A_{\alpha\beta\gamma}$ with a potential $V(A^{2})$ and a Dirac spinor field $\psi$ in a curve space-time can be constructed as
\begin{equation}\label{}
  \mathcal{L}_{m}=-\frac{1}{48}F^{2}-V(A^{2})+\frac{i}{2}[\bar{\psi}\Gamma^{\mu}D_{\mu}\psi-D_{\mu}\bar{\psi} \\ \Gamma^{\mu}\psi]-M(A^{2})\bar{\psi}\psi
\end{equation}
where $F=dA$ represents the field strength tensor and $D_{\mu}$ is the covariant derivative of spinor which satisfies
\begin{equation}\label{}
  D_{\mu}\psi = \partial_{\mu}\psi+\Omega_{\mu}\psi
\end{equation}
\begin{equation}\label{}
  D_{\mu}\bar{\psi} = \partial_{\mu}\bar{\psi}-\bar{\psi}\Omega_{\mu}
\end{equation}
The $\Omega_{\mu}=\frac{1}{2}\omega_{\mu a b}\Sigma^{a b}$ appeared in (2),(3) denotes the spin connection which is constituted by the Ricci spin coefficients $\omega_{\mu a b}=e_{a}^{\nu}\nabla_{\mu}e_{\nu b}$  and the generators of the spinor representation of the Lorentz group $\Sigma^{a b}=\frac{1}{4}[\gamma^{a},\gamma^{b}]$ . $\gamma^{a}$ and $\Gamma^{\mu}=e_{a}^{\mu}\gamma^{a}$ are the Dirac-Pauli matrices and their curve space-time counterparts respectively.
Following the general covariance principle, the tetrad $e_{a}^{\nu}$ is related to the metric by  $g^{\mu\nu}=e_{a}^{\mu}e_{b}^{\nu}\eta^{a b}$ with $\eta^{a b}=diag(1,-1,-1,-1)$. In such Lagrangians, the coupling between two fields is demonstrated by the the function $M(A^{2})$ which is the effective mass of dark matter.

One now can obtains the field equations from the total action
\begin{equation}\label{}
 S[A,g,\psi,\bar{\psi}]=\int\mathcal{L}\sqrt{-g}d^{4}x
\end{equation}
where $\mathcal{L}=\mathcal{L}_{g}+\mathcal{L}_{m}=\frac{R}{2\kappa^{2}}+\mathcal{L}_{m}$ is the Lagrangian including gravity,
$R$ denotes the Ricci scalar and $\kappa=\sqrt{8\pi G}$ is the inverse of the reduced Planck mass.

By varying the total action with respect to the three-form field and the Dirac field, we have the following equations of motion which are quite similar to that of electrodynamics
\begin{equation}\label{}
  \nabla_{\alpha}F^{\alpha\mu\nu\rho} = 12(\frac{dV}{dA^{2}}+\frac{dM}{dA^{2}}{\bar\psi \psi})A^{\mu\nu\rho}
\end{equation}
\begin{eqnarray}
  iD_{\alpha}\bar{\psi}\Gamma^{\alpha}+M\bar{\psi} &=& 0 \\
  -i\Gamma^{\alpha}D_{\alpha}\psi+M\psi &=& 0
\end{eqnarray}
The variation of the action with respect to the tetrad leads to Einstein field equation
\begin{equation}\label{}
  R_{\mu\nu}-\frac{1}{2}g_{\mu\nu}R=-\kappa^{2}T_{\mu\nu}
\end{equation}
where the total energy-momentum tensor for two fields is given by
\begin{equation}\label{}
\begin{split}
 T_{\mu\nu}=&-\frac{1}{6}F_{\mu\alpha\beta\gamma}F_{\nu}^{\alpha\beta\gamma}-6(\frac{dV}{dA^{2}}+\frac{dM}{dA^{2}}\bar\psi\psi)A_{\mu}^{\alpha\beta}A_{\nu\alpha\beta}\\
 &+\frac{1}{2}Re[\bar{\psi}i(\Gamma_{\mu}D_{\nu}+\Gamma_{\nu}D_{\mu})\psi]-g_{\mu\nu}\mathcal{L}_{m}.
\end{split}
\end{equation}
In the end of this section, we show the dual description of the above three-form field theory to place this field theory in a context more familiar to most of the community, such dual description has the following action:
\begin{equation}\label{}
 \widetilde{S}[\widetilde{A},g,\psi,\bar{\psi}]=\int\mathcal{\widetilde{L}}\sqrt{-g}d^{4}x
\end{equation}
where $\mathcal{\widetilde{L}}=\mathcal{L}_{g}+\mathcal{\widetilde{L}}_{m}$ and
\begin{equation}\label{}
  \mathcal{\widetilde{L}}_{m}=({\nabla_{\alpha}\widetilde{A}^{\alpha}})^2-\widetilde{V}(\widetilde{A}^{2})+\frac{i}{2}[\bar{\psi}\Gamma^{\mu}D_{\mu}\psi-D_{\mu}\bar{\psi} \\ \Gamma^{\mu}\psi]-\widetilde{M}(\widetilde{A}^{2})\bar{\psi}\psi
\end{equation}
$\widetilde{A}$ represents the dual of the three-form. $\widetilde{V}(\widetilde{A}^{2})$ and $\widetilde{M}(\widetilde{A}^{2})$ represent the self-coupling of $\widetilde{A}$ and the coupling between two fields respectively.
By varying the action with respect to the $\widetilde{A}$, $\psi$ and $\bar{\psi}$, we have the following equations of motion
\begin{equation}\label{}
  \partial_{\mu}\nabla_{\alpha}\widetilde{A}^{\alpha}+(\frac{d\widetilde{V}}{d\widetilde{A}^{2}}+\frac{d\widetilde{M}}{d\widetilde{A}^{2}}{\bar\psi \psi})\widetilde{A}_{\mu}= 0
\end{equation}
\begin{eqnarray}
  iD_{\alpha}\bar{\psi}\Gamma^{\alpha}+\widetilde{M}\bar{\psi} &=& 0 \\
  -i\Gamma^{\alpha}D_{\alpha}\psi+\widetilde{M}\psi &=& 0
\end{eqnarray}
which are different from that of three-form model.
\section{Cosmological evolution of the power-law coupled three-form dark energy model}
\label{sec:2}
We now consider the field equations in a homogeneous, isotropic, and spatially flat space-time described by the metric
\begin{equation}\label{}
  ds^{2}=dt^{2}-a(t)^{2}d\vec{x}^{2}
\end{equation}
where $a(t)$ refers to the scale factor.

To be compatible with FRW symmetries, the three-form field is assumed as the time-like component of the dual vector field, i.e.
\begin{equation}
  A_{i j k}=X(t)a(t)^{3}\varepsilon_{ijk}
\end{equation}
Since a three-form field without a potential can accelerates the expansion rate of the current universe\footnote{ A three-form field without a potential is equivalent to a cosmological constant\cite{Koivisto2009Three}.}, for simplicity, we set the potential to be zero in the following discussions, together with choosing the coupling function as the following power-law form
\begin{equation}\label{}
  M=m[(-\frac{\kappa^{2}}{6}A^{2})]^{\frac{\lambda}{2}}=m(\kappa \mid X \mid)^{\lambda}
\end{equation}
\footnote{ For simplicity, we consider $X \geq 0$ and neglect the absolute value sign in the following discussion.}
we have the Friedmann equations
\begin{eqnarray}
  H^{2} &=& \frac{\kappa^{2}}{3}\rho \\
  \dot{H} &=&-\frac{\kappa^{2}}{2}(\rho+p)
\end{eqnarray}
with
\begin{eqnarray}
  \rho &=& T_{0}^{0}=g^{00}T_{00}=\frac{1}{2}(3HX+\dot{X})^{2}+m(\kappa X)^{\lambda}\bar{\psi}\psi \\
  p &=& -T_{i}^{i}=-g^{ii}T_{ii}=-\frac{1}{2}(3HX+\dot{X})^{2}+\lambda m(\kappa X)^{\lambda}\bar{\psi}\psi
\end{eqnarray}
$\lambda$ is a dimensionless constant representing the strength of the coupling, this means that if $\lambda=0$, such field theory becomes a free field theory, so the constant $m$ with mass dimension is, in fact, the mass of dark matter in a free field theory. Since typically it is very difficult for dark energy to couple dark matter with mass bigger than milli-eV \cite{d2016quantum}, we choose $m$ to be smaller than milli-eV. Although the phenomenological bounds on dark matter mass coming from large scale structure require that most of dark matter is considerably heavier than $10^{-3} $ eV \cite{Dodelson1994Sterile}, some dark matter's mass can be smaller than milli-eV. Indeed, Rajagopal, Turner and Wilczek considered axino in the keV range and they obtained the axino mass bound $m_{a}<2$ keV for axino to be warm dark matter\cite{Rajagopal1990Cosmological}.

In the FRW space-time, there is only one independent equation of motion of the three-form field
\begin{equation}\label{}
  \ddot{X}+3(\dot{H}X+H\dot{X})+\lambda\kappa m(\kappa X)^{\lambda-1}\bar{\psi}\psi=0
\end{equation}
 from the equations of motion of the spinor field and its Dirac adjoint, one can obtains the following equation
\begin{equation}\label{}
  \frac{d(\bar{\psi}\psi)}{dt}+3H(\bar{\psi}\psi)=0
\end{equation}
with its simple solution
\begin{equation}\label{}
  \bar{\psi}\psi=(\bar{\psi}\psi)_{0}a^{-3}
\end{equation}
which shows that our model indeed returns to $\Lambda$CDM model when the coupling constant $\lambda$ becomes $0$.

Providing with the equations of motion, we have the continuity equations for both components
\begin{eqnarray}
  \dot{\rho}_{X}+3H(\rho_{X}+p_{X}) &=& -\delta H\rho_{m} \\
  \dot{\rho}_{m}+3H(\rho_{m}+p_{m}) &=& \delta H\rho_{m}
\end{eqnarray}
with
\begin{eqnarray}
  \rho_{X} &=& \frac{1}{2}(3HX+\dot{X})^{2}, \hspace{1cm}   p_{X}=p \\
  \rho_{m} &=& m(\kappa X)^{\lambda}\bar{\psi}\psi, \hspace{1cm}   p_{m}=0
\end{eqnarray}
\begin{equation}\label{}
\delta=\lambda\frac{X^{\prime}}{X}
\end{equation}
the prime stands for derivative with respect to e-folding time $N=\ln a$ here and in the following.

In order to study cosmological dynamics in such coupled dark energy model, it is convenient to introduce the following dimensionless variable\cite{Koivisto2009Inflation2}
\begin{equation}\label{}
  x=\kappa X, \hspace{1cm} y=\frac{\kappa}{\sqrt{6}}(X^{\prime}+3X), \hspace{1cm} \omega^{2}=\frac{\kappa^{2}\rho_{m}}{3H^{2}}.
\end{equation}
By applying the Friedmann equations and equations of motion, one can obtains the autonomous system of evolution equations
\begin{eqnarray}
  x^{\prime} &=& \sqrt{6}y-3x\\
  y^{\prime} &=& \frac{1}{2}\left(\frac{\lambda}{x}\left(\sqrt{6}-3xy\right)-3y\right)\left(y^{2}-1\right).
\end{eqnarray}
We note that $\omega^{2}$ has been eliminated by the Friedmann constraint written in terms of these variables $y^{2}+\omega^{2}=1$.

There are two fixed points for such autonomous system. One of them is $\left(\sqrt{\frac{2}{3}},1\right)$, which is an attractor since its eigenvalues $(-3,-3)$ are both negative. By rewriting density and pressure in term of the dimensionless variables, we have the total EOS
\begin{equation}\label{}
  \omega_{tot}=\frac{p}{\rho}=-1+(1+\lambda)\left(1-y^{2}\right)=-1
\end{equation}
indicating that such fixed point represents a three-form saturated de Sitter universe.
The other one $\left(\sqrt{\frac{2\lambda}{3(1+\lambda)}},\sqrt{\frac{\lambda}{(1+\lambda)}}\right)$ with eigenvalues $\frac{3}{4}\left(\left(-1+\sqrt{17}\right),\left(-1-\sqrt{17}\right)\right)$
is a saddle point, strictly this fixed point exists only when $\lambda>0$.
It can be inferred from $\omega_{X}=\frac{p_{X}}{\rho_{X}}=-1+\frac{\lambda\left(1-y^{2}\right)}{y^{2}}=0$, $\delta=0$ that such fixed point is a tracking solution which can be used to alleviate the coincidence problem with a fine-turning of the model parameters.

\begin{figure}
  \includegraphics[width=0.55\textwidth]{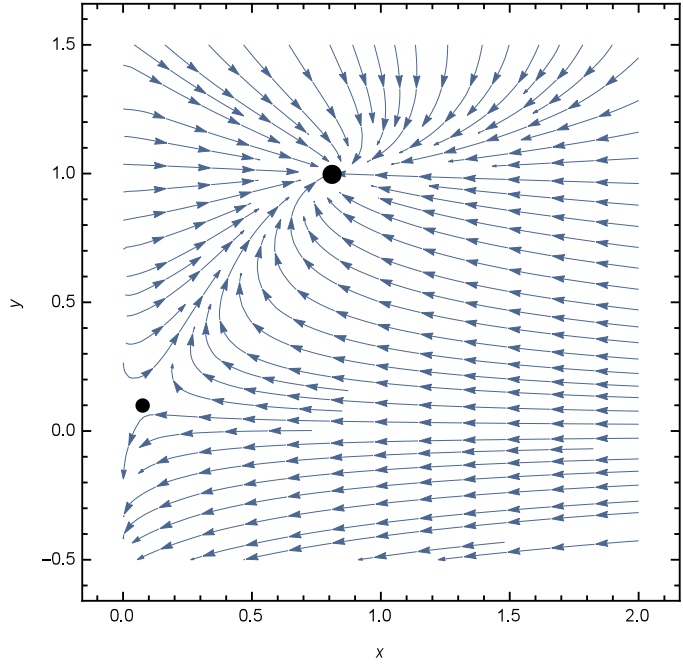}
\caption{The larger black point represents the stable point,and the smaller one represents the saddle point.}
\label{fig:1}       
\end{figure}

The trajectories with respect to $x(N)$ and $y(N)$ with a wide range of initial conditions and a assumption that $\lambda=0.01$ (we will see that this is a good choice in the next section) can be visualized by Fig.\ref{fig:1}.

As is showed in the Fig.\ref{fig:1}, the trajectories run toward the de Sitter attractor, coasting along the saddle point. Also, one should note that the present value of $x$ must be equal to or larger than a certain value $x_{*}$ that depends on both $\lambda$ and $y(0)$ to make sure in high redshift the autonomous system does not encounter singularity. After specifying $\lambda$ and $y(0)$, if $\widetilde{x}$ is the present value of $x$ leading to that $x(N)$ is positive-definite for arbitrary non-infinite $N$, i.e. $x(N)>0(-\infty <N< +\infty)$, then $x_{*}$ is the lower limit of $\widetilde{x}$.

Now let us solve the the autonomous system of evolution equations by assuming a large $x_{0}$. With the constraint $y^{2}<1$, it can be well approximated by two independent equations
\begin{eqnarray}
  x^{\prime} &=& -3x \\
  y^{\prime} &=& \frac{3}{2}(1+\lambda)y(1-y^{2})
\end{eqnarray}
which have the following solutions
\begin{eqnarray}
  x & \approx & x_{0}(1+z)^{3} \\
  y^{2} & \approx & \frac{1}{1+\frac{\Omega_{m0}}{1-\Omega_{m0}}(1+z)^{3(1+\lambda)}}
\end{eqnarray}
we have replaced e-folding time by the redshift here.

 Substituting the solutions into(27)-(29), we have
\begin{eqnarray}
  \rho_{X} &\approx & \frac{3H_{0}^{2}(1-\Omega_{m0})}{\kappa^{2}} \\
  p_{X} &\approx & -\frac{3H_{0}^{2}(1-\Omega_{m0})}{\kappa^{2}}+\lambda\frac{3H_{0}^{2}\Omega_{m0}}{\kappa^{2}}(1+z)^{3(1+\lambda)}\\
  \delta &\approx & -3\lambda
\end{eqnarray}
noting that although $\delta$ is a constant, such solutions are different from the models proposed in\cite{Wang2004Can,wei2007observational,amendola2007consequences} since
\begin{equation}\label{}
  \omega_{X}\approx -1+\lambda\frac{\Omega_{m0}}{1-\Omega_{m0}}(1+z)^{3(1+\lambda)}
\end{equation}
is not a constant.

It can be inferred from (40) and (41) that the energy transfer between two dark sectors keeps the density of dark energy as a constant even when its EOS deviates from $-1$.

To compare the solutions with uncoupled dark energy model, let us rewrite Hubble parameter as
\begin{equation}\label{}
\begin{split}
   \frac{H^{2}(z)}{H_{0}^{2}} &=\frac{\Omega_{m0}\left(\frac{x}{x_{0}}\right)^{\lambda}(1+z)^{3}}{1-y^{2}}  \\
     &=\Omega_{m0}(1+z)^{3}+(1-\Omega_{m0})\frac{\rho_{Xeff}}{\rho_{Xeff0}}
\end{split}
\end{equation}
where
\begin{equation}\label{}
\begin{split}
\frac{\rho_{Xeff}}{\rho_{Xeff0}} &=\frac{\Omega_{m0}}{1-\Omega_{m0}}\left(\frac{\left(\frac{x}{x_{0}}\right)^{\lambda}(1+z)^{3}}{1-y^{2}}-(1+z)^{3}\right)\\
&=\exp\left[\int_{0}^{z}\frac{3(1+\omega_{Xeff}(\tilde{z}))}{1+\tilde{z}}d\tilde{z}\right]
\end{split}
\end{equation}

is the normalized effective dark energy density and
\begin{equation}\label{}
  \omega_{Xeff}(z)=-1+\frac{1}{3}\frac{x^{\lambda}\left(3+\frac{\lambda(1+z)}{x}\frac{dx}{dz}\right)(1-y^{2})+(1+z)\frac{dy^{2}}{dz}x^{\lambda}-3(1-y^{2})^{2}x_{0}^{\lambda}}{\left(x^{\lambda}-(1-y^{2})x_{0}^{\lambda}\right)(1-y^{2})}.
\end{equation}
is the effective EOS of dark energy.

By substituting the approximate solutions into (43) and (44), we have
\begin{eqnarray}
  \frac{\rho_{Xeff}}{\rho_{Xeff0}} &\approx & 1+\frac{\Omega_{m0}}{1-\Omega_{m0}}\left((1+z)^{3(1+\lambda)}-(1+z)^{3}\right) \\
  \omega_{Xeff} &\approx& \frac{-1+\lambda\frac{\Omega_{m0}}{1-\Omega_{m0}}(1+z)^{3(1+\lambda)}}{1+\frac{\Omega_{m0}}{1-\Omega_{m0}}\left((1+z)^{3(1+\lambda)}-(1+z)^{3}\right)}
\end{eqnarray}
as one can see, depending on the sign of $\lambda$, there are two different high redshift approximate expressions for the effective energy density or the effective EOS. More specifically, effective energy density is approximate to $\frac{\Omega_{m0}}{1-\Omega_{m0}}(1+z)^{3(1+\lambda)}(\lambda >0)$ and $-\frac{\Omega_{m0}}{1-\Omega_{m0}}(1+z)^{3}(\lambda <0)$ at high redshift, and the effective EOS of dark energy is approximate to $\lambda(\lambda >0)$ and $\lambda(1+z)^{3\lambda}(\lambda <0)$ at high redshift.

At the end of this section, we place another restriction on the likelihood function of $(x_{0},\lambda,\Omega_{m0})$ with the aid of the approximate solution of the autonomous system of evolution equations, in fact we can see from (45) and (46) immediately that the Hubble parameter is independent of $x_{0}$ supposing $x_{0}$ take a large value, which means that the likelihood function becomes a none zero constant(in fact a large $x_{0}$ with proper values of $\lambda$ and $\Omega_{m0}$ is quite favored by observations) with respect to $x_{0}$ if $x_{0}$ is large enough.
Given this behavior of the likelihood function $L$, we have such formula
\begin{equation}\label{}
  \int_{0}^{+\infty}L(x_{0},\lambda,\Omega_{m0})dx_{0} \propto \lim_{x_{0}\rightarrow+\infty}L(x_{0},\lambda,\Omega_{m0})
\end{equation}
which can be used to marginalize over $x_{0}$ without a prior.

\section{Confront the power-law coupled three-form dark energy model with observations}
\label{sec:3}
In this section, we perform a simple likelihood analysis on the free parameters of the model with the combination of data from Type \uppercase\expandafter{\romannumeral1}a Supernova (SN \uppercase\expandafter{\romannumeral1}a), Baryon Acoustic Oscillations (BAO) and Cosmic Microware Backround (CMB) radiation observations.

Firstly, we construct the following $ \chi^{2}$ function for SN \uppercase\expandafter{\romannumeral1}a by using the recently released Union2.1 compilation with 580 data points
\begin{equation}\label{}
  \chi_{SN \uppercase\expandafter{\romannumeral1}a}^{2}= P- \frac{Q^{2}}{R}
\end{equation}
where $P$, $Q$ and $R$ are defined as
\begin{eqnarray}
  P &=& \sum_{i=0}^{580}\frac{\left(\mu_{th}(z_{i})-\mu_{obs}(z_{i})\right)^{2}}{\sigma_{\mu}^{2}(z_{i})},\\
  Q &=& \sum_{i=0}^{580}\frac{\left(\mu_{th}(z_{i})-\mu_{obs}(z_{i})\right)}{\sigma_{\mu}^{2}(z_{i})}, \\
  R &=& \sum_{i=0}^{580}\frac{1}{\sigma_{\mu}^{2}(z_{i})},
\end{eqnarray}
with $\mu_{th}=5\log_{10}\left[(1+z)\int_{0}^{z}\frac{H_{0}}{H(z^{\prime})}dz^{\prime}\right]+25$ denotes the distance modulus predicted by theory and $\mu_{obs}$ represents the observed one with a statistical uncertainty  $\sigma_{\mu}$.

\begin{figure}[h]
\begin{minipage}{0.48\linewidth}
  \centerline{\includegraphics[width=1\textwidth]{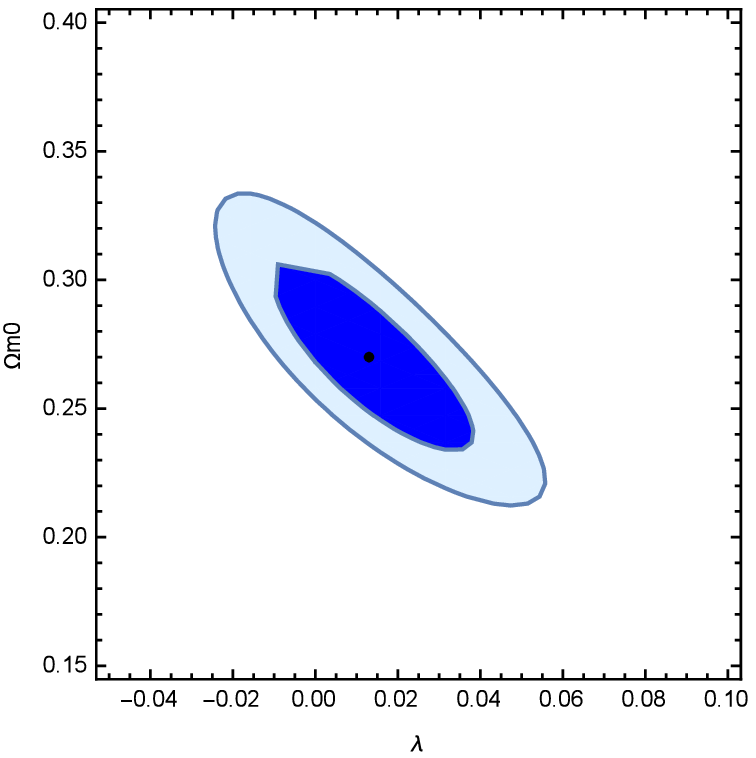}}
  \label{a}
\end{minipage}
\hfill
\begin{minipage}{.48\linewidth}
  \centerline{\includegraphics[width=1\textwidth]{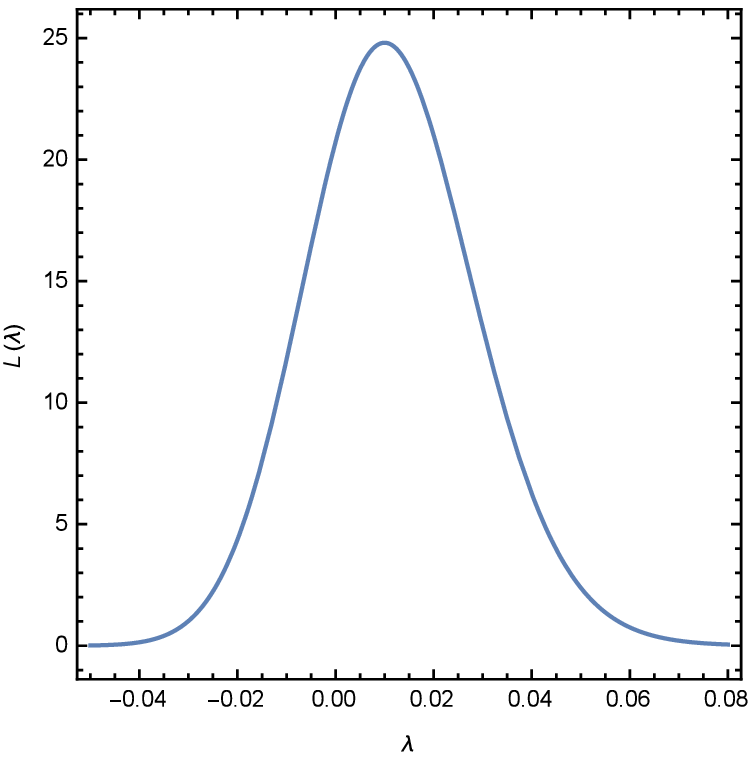}}
\end{minipage}
\caption{The figure on the left shows observational constraints on parameters $(\lambda,\Omega_{m0})$ with the combination of SN \uppercase\expandafter{\romannumeral1}a and BAO/CMB detasets, in which the light blue and blue region are corresponding to $2\sigma$  and $1\sigma$ region respectively, while the black point (0.013,0.269) with $\chi^{2}=564.811$ represents the best-fit value of the pair $(\lambda,\Omega_{m0})$. The figure on the right is the likelihood function of $\lambda$ which is marginalized with a flat prior that becomes zero if $\Omega_{m0}$ bigger than 0.38 or smaller than 0.16, suggesting $-0.017<\lambda<0.047$ ($2\sigma$ confidence level). }
\label{fig:2}
\end{figure}

In the second step, we consider BAO data from the WiggleZ Survey, SDSS DR7 Galaxy sample and 6dF Galaxy Survey together with CMB data from WMAP 7 yeas observations to obtain the BAO/CMB constraints on the model parameters by defining $\chi_{BAO/CMB}^{2}$ as\cite{Giostri2012From,Mamon2016Constraints,Mamon2016A}

\begin{equation}\label{}
  \chi_{BAO/CMB}^{2}=X^{T}C^{-1}X
\end{equation}
where
\begin{equation}\label{}
X=\left(
\begin{matrix}
\frac{d_{A}(z_{*})}{D_{v}(0.106)}-30.95\\
\frac{d_{A}(z_{*})}{D_{v}(0.2)}-17.55\\
\frac{d_{A}(z_{*})}{D_{v}(0.35)}-10.11\\
\frac{d_{A}(z_{*})}{D_{v}(0.44)}-8.44\\
\frac{d_{A}(z_{*})}{D_{v}(0.6)}-6.69\\
\frac{d_{A}(z_{*})}{D_{v}(0.73)}-5.45
\end{matrix}
\right)
\end{equation}
in which $d_{A}(z)=\int_{0}^{z}\frac{1}{H(z^{\prime})}dz^{\prime}$ and $D_{V}(z)=\left[d_{A}(z)^{2}\frac{z}{H(z)}\right]^{\frac{1}{3}}$ represent the co-moving angular-diameter distance and the dilation scale respectively, while $z_{*}\approx1091$ is the decoupling time.
\begin{equation*}
C^{-1}=\left(
\begin{matrix}
  0.48435 & -0.101383 & -0.164945 & -0.0305703 & -0.097874 & -0.106738 \\
  -0.101383 & 3.2882 & -2.45497 & -0.0787898 & -0.252254 & -0.2751 \\
  -0.164945 & -2.45497 & 9.55916 & -0.128187 & -0.410404 & -0.447574 \\
  -0.0305703 & -0.0787898 & -0.128187 & 2.78728 & -2.75632 & 1.16437 \\
  -0.097874 & -0.252254 & -0.410404 & -2.75632 & 14.9245 & -7.32441 \\
  -0.106738 & -0.2751 & -0.447574 & 1.16437 & -7.32441 & 14.5022
\end{matrix}
\right)
\end{equation*}
is the inverse of the correlation matrix.

Finally, the total $\chi^{2}$ function for the combined observational datasets is given by $\chi^{2}=\chi_{SN \uppercase\expandafter{\romannumeral1}a}^{2}+\chi_{BAO/CMA}^{2}$, from which we can construct the likelihood function as $L=L_{0} e^{-\frac{1}{2}\chi^{2}}$, here $L_{0}$ is a normalized constant which is independent of the free parameters.

Providing with the likelihood function, one can then obtain the best-fit values of the free parameters by maximizing it. However, as was mentioned in the previous section, the parameter $x_{0}$ can't be strictly restricted, so we leave it out of the discussions and consider the likelihood function that has been marginalized over $x_{0}$ without a prior. By the aid of (47), such function is proportional to
 \begin{equation}\label{}
 \lim_{x_{0}\rightarrow+\infty}L(x_{0},\lambda,\Omega_{m0})\approx L(\tilde{x_{0}},\lambda,\Omega_{m0})
\end{equation}
 where $\tilde{x_{0}}$ is a large number, one can choose it as 10000, for example. We now present the fitting result in Fig.\ref{fig:2} by analyzing such likelihood.

One may note that the marginalized likelihood in the right panel of Fig.\ref{fig:2} appears a little non-Gaussian, this is mainly because of the non-Gaussian structure of the likelihood that haven't been marginalized. One also can see from the Fig.\ref{fig:2} that observations favor a small positive coupling constant which, as we mentioned above, allows the existence of a tracking solution that can be used to alleviate the coincidence problem with a fine-turning of the model parameters.

\begin{figure*}
  \includegraphics[width=0.75\textwidth]{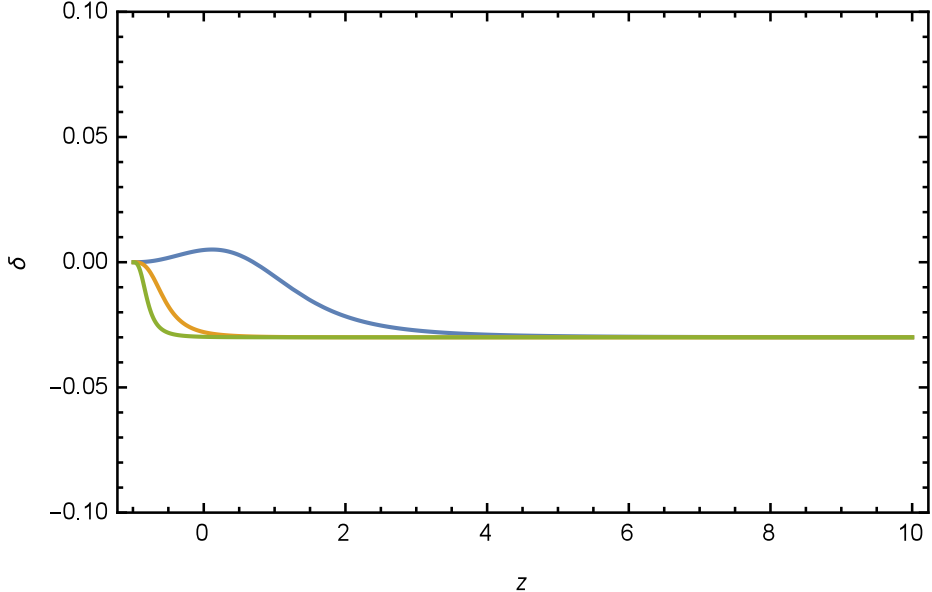}
\caption{$\delta=\lambda\frac{x^{\prime}}{x}$ here represents the strength of the interactions, the blue, orange and green curve are corresponding to the case with $(\lambda,x_{0},\Omega_{m0})=(0.01,0.6,0.27),(0.01,10,0.27),(0.01,100,0.27)$ respectively.}
\label{fig:3}       
\end{figure*}

Moreover, one thing here needs to be noticed, since $x_{0}$ can't be strictly restricted, the interact behavior between two dark sectors still remains uncertain, which can be inferred from Fig.\ref{fig:3}. However, one may decrease such uncertainty by taking into account observational constraints from future measurements.

\begin{figure}
  \includegraphics[width=0.75\textwidth]{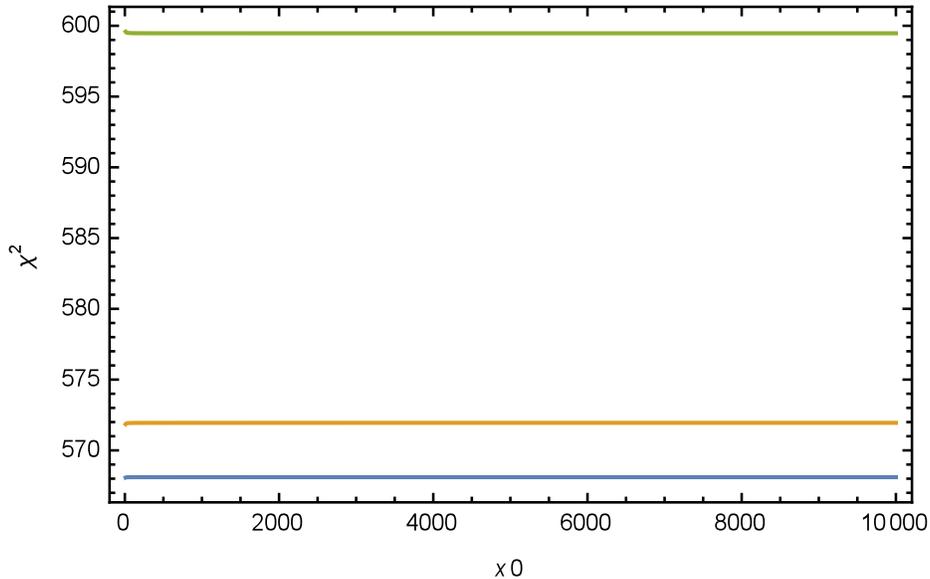}
\caption{The blue, orange and green line are corresponding to the case with $(\lambda,\Omega_{m0})=(0.01,0.3),(0.02,0.3),(0.01,0.2)$ respectively. Here $x_{0}$ ranges from 10 to 10000.}
\label{fig:4}       
\end{figure}

From Fig.\ref{fig:3}, it can be inferred that the direction of energy transfer can be changed if $x_{0}$ is sufficiently close to $x_{*}$, which is around 0.6 in such case, and the behavior of $\delta$ is almost the same between the choices of $x_{0}=10$ and $x_{0}=100$ if the redshift $z>0$, which is consistent with the conclusion that the likelihood function becomes a constant with respect to $x_{0}$ if $x_{0}$ is adequately large that we have drawn in the section \ref{sec:2}. In fact we can prove this conclusion in a inductively way by plotting $\chi^{2}$ (see Fig.\ref{fig:4}).

\section{Conclusions}
\label{sec:4}

In this paper we have studied a power-law coupled dark energy model which considers dark
energy as a three-form field and dark matter as a spinor field. By performing a dynamical analysis on the field
equations with the introduction of three dimensionless variables, we
obtained two fixed points of the autonomous system of evolution equations, among which one is a de Sitter attractor, and the other is a tracking solution, supposing $\lambda>0$, that provides a possible solution of the coincidence problem.

By marginalizing over $x_{0}$, we have also carried out a likelihood analysis on the free parameters $\lambda$ and $\Omega_{m0}$ with the combination of SN $\uppercase\expandafter{\romannumeral1}$a+BAO/CMB datasets, through which we have a best-fit value of the pair $(\lambda,\Omega_{m0})$ as $(0.013,0.269)$.
In addition, the likelihood function marginalized over $x_{0}$ and $\Omega_{m0}$ showed that $\lambda$ is restricted by $-0.017<\lambda<0.047$ ($2\sigma$ confidence level, with a best-fit value $0.01$), indicating that the measurements considered here are quite consistent between $\Lambda CDM$ and our three-form model. However, future measurements might allow us to tell them apart.

Notwithstanding it can be told from the fitting result that $\lambda$ and $\Omega_{m0}$ are strictly restricted, $x_{0}$ can be any value beyond $x_{*}$. However, as mentioned above, future measurements might decrease the uncertainty on $x_{0}$.

\section*{Acknowledgments}
The authors warmly thank Jia-Xin Wang and Deng Wang for beneficial discussions.

\bibliographystyle{spphys}
\bibliography{coupled}

\begin{thebibliography}{10}
\providecommand{\url}[1]{{#1}}
\providecommand{\urlprefix}{URL }
\expandafter\ifx\csname urlstyle\endcsname\relax
  \providecommand{\doi}[1]{DOI \discretionary{}{}{}#1}\else
  \providecommand{\doi}{DOI \discretionary{}{}{}\begingroup
  \urlstyle{rm}\Url}\fi

\bibitem{riess1998}
A.G. Riess, A.V. Filippenko, P.~Challis, A.~Clocchiatti, A.~Diercks, P.M.
  Garnavich, R.L. Gilliland, C.J. Hogan, S.~Jha, R.P. Kirshner, et~al., The
  Astronomical Journal \textbf{116}(3), 1009 (1998)

\bibitem{perlmutter1999measurements}
S.~Perlmutter, G.~Aldering, G.~Goldhaber, R.~Knop, P.~Nugent, P.~Castro,
  S.~Deustua, S.~Fabbro, A.~Goobar, D.~Groom, et~al., The Astrophysical Journal
  \textbf{517}(2), 565 (1999)

\bibitem{Sahni2004Dark}
V.~Sahni, Lecture Notes in Physics \textbf{653}(2), 141 (2004)

\bibitem{Carroll2000The}
S.M. Carroll, Living Reviews in Relativity \textbf{4}(1), 1 (2001)

\bibitem{Caldwell1998Cosmological}
R.R. Caldwell, R.~Dave, P.J. Steinhardt, Physical Review Letters
  \textbf{80}(8), 1582 (1998)

\bibitem{Caldwell1999A}
R.R. Caldwell, Phys.lett \textbf{45}(3), 549 (1999)

\bibitem{Piazza2004Dilatonic}
F.~Piazza, S.~Tsujikawa, Journal of Cosmology and Astroparticle Physics
  \textbf{2004}(07), 004 (2004)

\bibitem{Padmanabhan2002Accelerated}
T.~Padmanabhan, Physical Review D Particles Fields \textbf{66}(2), 611 (2002)

\bibitem{Bo2006Oscillating}
B.~Feng, M.~Li, Y.S. Piao, X.~Zhang, Physics Letters B \textbf{634}(2), 101
  (2006)

\bibitem{Koivisto2009Inflation1}
T.S. Koivisto, D.F. Mota, C.~Pitrou, Journal of High Energy Physics
  \textbf{2009}(09), 092 (2009)

\bibitem{Koivisto2009Three}
T.S. Koivisto, N.J. Nunes, Physics Letters B \textbf{685}(2), 105 (2010)

\bibitem{mulryne2012three}
D.J. Mulryne, J.~Noller, N.J. Nunes, Journal of Cosmology and Astroparticle
  Physics \textbf{2012}(12), 016 (2012)

\bibitem{Koivisto2012Coupled}
T.S. Koivisto, N.J. Nunes, Physical Review D \textbf{88}(12), 123512 (2013)

\bibitem{Amendola1999Coupled}
L.~Amendola, Physical Review D \textbf{62}(4), 043511 (2000)

\bibitem{Ngampitipan2011Dynamics}
T.~Padmanabhan, Physical Review D \textbf{66}(2), 021301 (2002)

\bibitem{Micheletti2009Field}
S.~Micheletti, E.~Abdalla, B.~Wang, Physical Review D \textbf{79}(12), 123506
  (2009)

\bibitem{Suzuki2012The}
N.~Suzuki, D.~Rubin, C.~Lidman, G.~Aldering, R.~Amanullah, K.~Barbary,
  L.~Barrientos, J.~Botyanszki, M.~Brodwin, N.~Connolly, et~al., The
  Astrophysical Journal \textbf{746}(1), 85 (2012)

\bibitem{Chris2011The}
C.~Blake, E.A. Kazin, F.~Beutler, T.M. Davis, D.~Parkinson, S.~Brough,
  M.~Colless, C.~Contreras, W.~Couch, S.~Croom, et~al., Monthly Notices of the
  Royal Astronomical Society \textbf{418}(3), 1707 (2011)

\bibitem{Percival2010Baryon}
W.J. Percival, B.A. Reid, D.J. Eisenstein, N.A. Bahcall, T.~Budavari, J.A.
  Frieman, M.~Fukugita, J.E. Gunn, {\v{Z}}.~Ivezi{\'c}, G.R. Knapp, et~al.,
  Monthly Notices of the Royal Astronomical Society \textbf{401}(4), 2148
  (2010)

\bibitem{Beutler2011The}
F.~Beutler, C.~Blake, M.~Colless, D.H. Jones, L.~Staveley-Smith, L.~Campbell,
  Q.~Parker, W.~Saunders, F.~Watson, Monthly Notices of the Royal Astronomical
  Society \textbf{416}(4), 3017 (2011)

\bibitem{jarosik2011seven}
N.~Jarosik, C.~Bennett, J.~Dunkley, B.~Gold, M.~Greason, M.~Halpern, R.~Hill,
  G.~Hinshaw, A.~Kogut, E.~Komatsu, et~al., The Astrophysical Journal
  Supplement Series \textbf{192}(2), 14 (2011)

\bibitem{Weinberg1973Gravitation}
S.~Weinberg, \emph{Gravitation and cosmology: principles and applications of
  the general theory of relativity} (Wiley New York, 1972)

\bibitem{birrell1984quantum}
N.D. Birrell, P.C.W. Davies, \emph{Quantum fields in curved space} (Cambridge
  university press, 1982)

\bibitem{wald1984general}
R.M. Wald, \emph{General Relativity} (The University of Chicago Press, 1984)

\bibitem{ryder1996quantum}
L.H. Ryder, \emph{Quantum field theory} (Cambridge university press, 1996)

\bibitem{d2016quantum}
G.~D¡¯Amico, T.~Hamill, N.~Kaloper, Physical Review D \textbf{94}(10), 103526
  (2016)

\bibitem{Dodelson1994Sterile}
D.~S, W.~LM, Physical review letters \textbf{72}(1), 17 (1994)

\bibitem{Rajagopal1990Cosmological}
K.~Rajagopal, M.S. Turner, F.~Wilczek, in \emph{International Symposium on
  Power Semiconductor Devices and ICS} (1990), pp. 254--257

\bibitem{Koivisto2009Inflation2}
T.S. Koivisto, N.J. Nunes, Physical Review D \textbf{80}(10), 103509 (2009)

\bibitem{Wang2004Can}
P.~Wang, X.H. Meng, Classical and Quantum Gravity \textbf{22}(2), 283 (2004)

\bibitem{wei2007observational}
H.~Wei, S.N. Zhang, Physics Letters B \textbf{644}(1), 7 (2007)

\bibitem{amendola2007consequences}
L.~Amendola, G.C. Campos, R.~Rosenfeld, Physical Review D \textbf{75}(8),
  083506 (2007)

\bibitem{Giostri2012From}
R.~Giostri, M.V. dos Santos, I.~Waga, R.~Reis, M.~Calvao, B.~Lago, Journal of
  Cosmology and Astroparticle Physics \textbf{2012}(03), 027 (2012)

\bibitem{Mamon2016Constraints}
A.A. Mamon, K.~Bamba, S.~Das, European Physical Journal C \textbf{77}(1), 29
  (2017)

\bibitem{Mamon2016A}
A.A. Mamon, S.~Das, International Journal of Modern Physics D \textbf{25}(03),
  1650032 (2016)

\end{thebibliography}

\end{document}